# Spin foam with topologically encoded tetrad on trivalent spin networks

**Raymond Aschheim**

Polytopics, 8 villa Haussmann, 92130 Issy, France

Raymond@Aschheim.com

**Abstract**. We explore discrete approaches in LQG where all fields, the gravitational tetrad, and the matter and energy fields, are encoded implicitly in a graph instead of being additional data. Our graph should therefore be richer than a simple simplicial decomposition. It has to embed geometrical information and the standard model. We start from Lisi's model. We build a trivalent graph which is an F4 lattice of 48-valent supernodes, reduced as trivalent subgraphs, and topologically encoding data. We show it is a solution for EFE with no matter. We define bosons and half-fermions in two dual basis. They are encoded by bit exchange in supernodes, operated by Pachner 2-2 move, and rest state can be restored thanks to information redundancy. Despite its 4 dimensional nature, our graph is a trivalent spin network, and its history is a pentavalent spin foam.

## 1. Motivation

Could a theory of everything be nothing more then the set theory ? Ultimate nature of Nature would be a set of cardinal three subsets. Shape of Nature comes from a unique, optimally symmetric geometric object: the sixth platonic element, the icositetrachoron (or 24-cell regular polytope), crystallized as a hyperdiamond network, mathematically and physically made of loops and connections.
This paper presents an approach to embed extended[1] standard model in the topology of a spin foam; where spacetime, geometry, matter and forces are emergent information from a simple trivalent graph.

## 2. Trivalent spin network

Our spin network is trivalent, but holds the (internal graph distance) metric of a regular lattice, that can be build 2D or 3D as toy models, and 4D in serious Hyperdiamond model discussed in 4 &5.

### 2.1. Regular spin network based on space filling

As illustrated for 2D (figure 1), a regular lattice can be triangulated to fill an infinite plan or a two dimensional torus (figure 2).

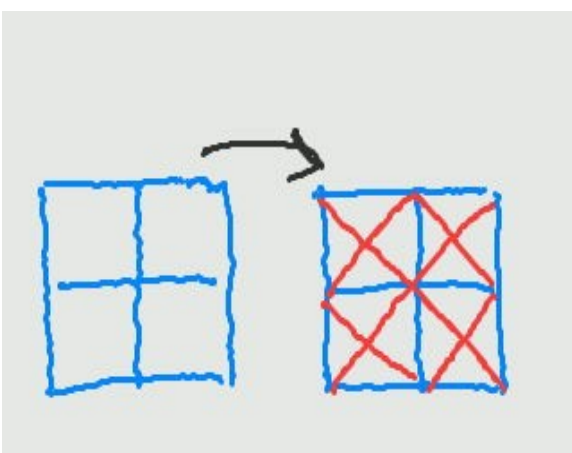

**Figure 1.** Triangulation of a square grid

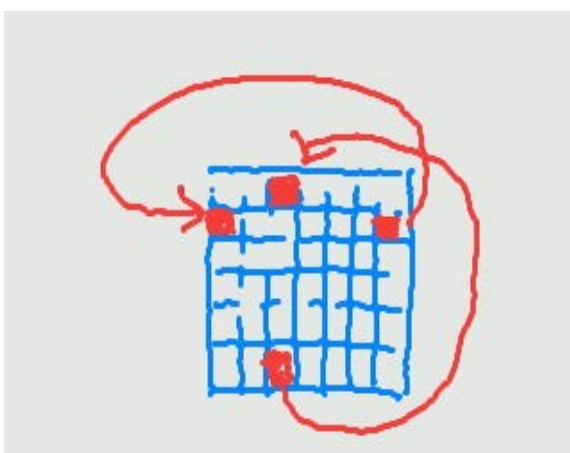

**Figure 2.** Toroidal topology, identification of first and last row/col

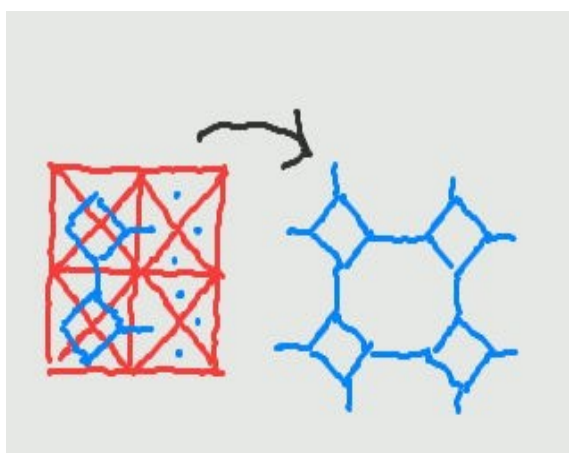

**Figure 3.** Dual two-complex network linking triangle centers.

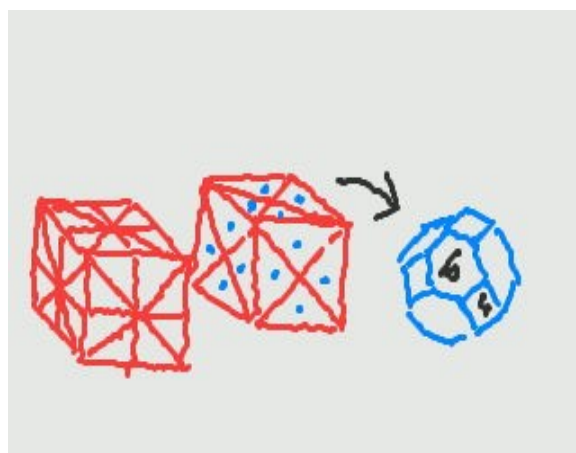

**Figure 4.** 3D cubic lattice & cantitruncated dual cubic honeycomb

*2.1.1. Two-complex dual of a triangulation.* The standard way to build spin network, taking the dual of a triangulation, gives trivalent spin network in 2D (figure 3), and n+1 valent in nD (figure 4).

*2.1.2. Valence reduction*. Spin network based on dual 2-complex of triangulation has valence = dimension + 1. It reduces to a trivalent by replacing any n-valent node by a n-valent supernode which is a part of a trivalent graph (in figure 5, a 4-valent node is replaced by two trivalent nodes), giving a trivalent spin network embedded in a 3D space $T^3$ discretized as cubic grid (figure 6). A trivalent spin network embedded in a 4D space $T^4$ discretized as hyperdiamond F4 lattice is better (figure 7 showing a slice of it with 48-valent supernodes). The number of supernodes in a $T^4$ of size 2n is $2n^4$.

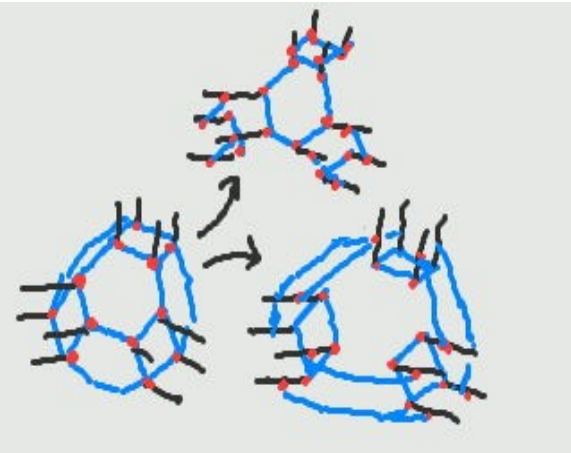

**Figure 5.** Quadrivalent nodes expanded into two trivalent

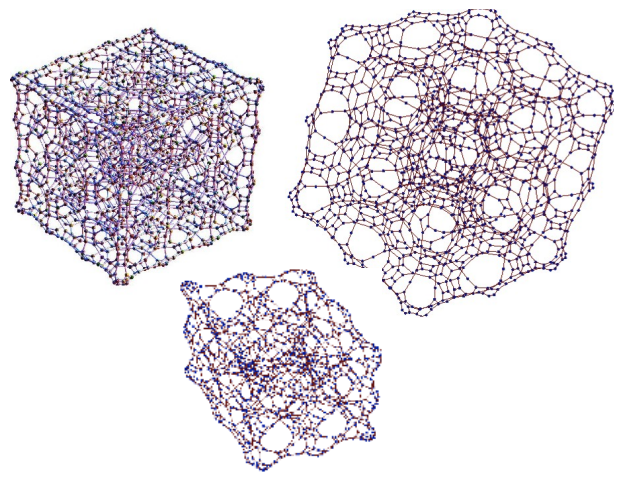

**Figure 6.** Trivalent tridimensional lattice

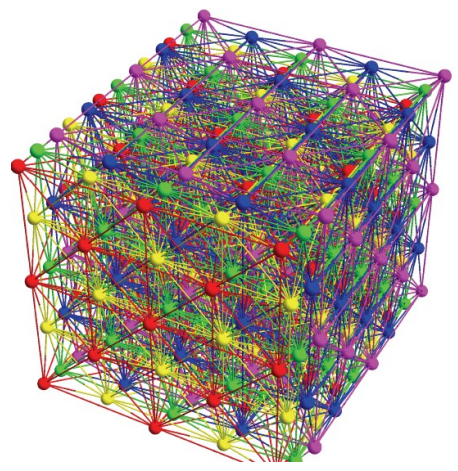

**Figure 7.** F4 lattice of 48valent nodes, sliced

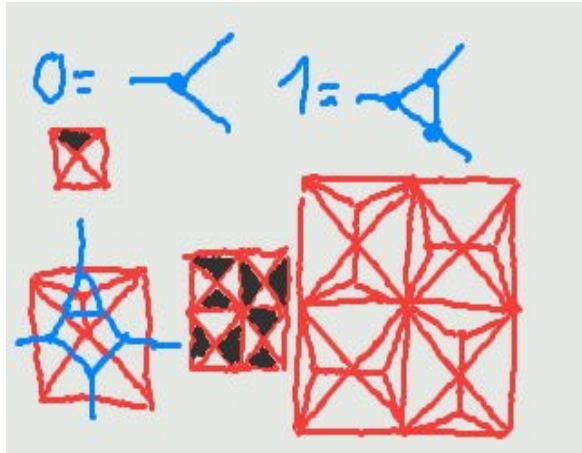

**Figure 8.** Bit encoding and checkerboard state

*2.1.3. Encoding bits*. Any node (figure 8) holds bit 1 if in a 3-loop, 0 otherwise. The 1 is a triangulation refinement. Checkerboard polarization is induced by alternating 0 and 1.

2.2. Trivalent Spin Network

**Informed regular trivalent** spin network gives a "Graphiton Model".

It has been proved [2] that bit defects induces gravitational field by curving topological distance graph geodesics (in a specific 2D case, figure 8), as an emerging property of a pure topological graph.

## 3. Spin foam

Spin foam is the history of an informed regular trivalent spin network, as illustrated in figure 9.

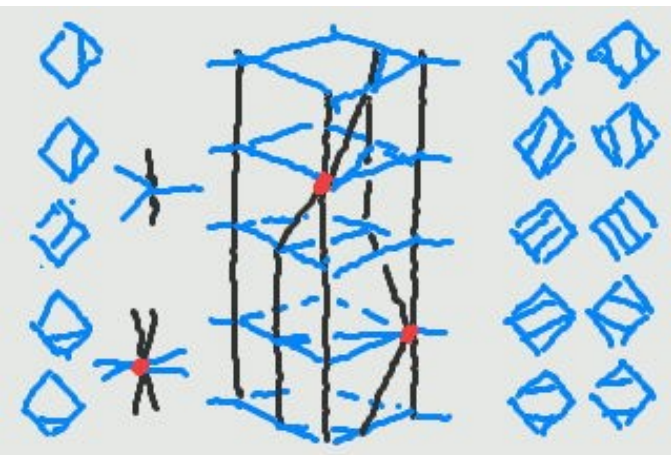

**Figure 9.** Spin foam history of a bit inversion

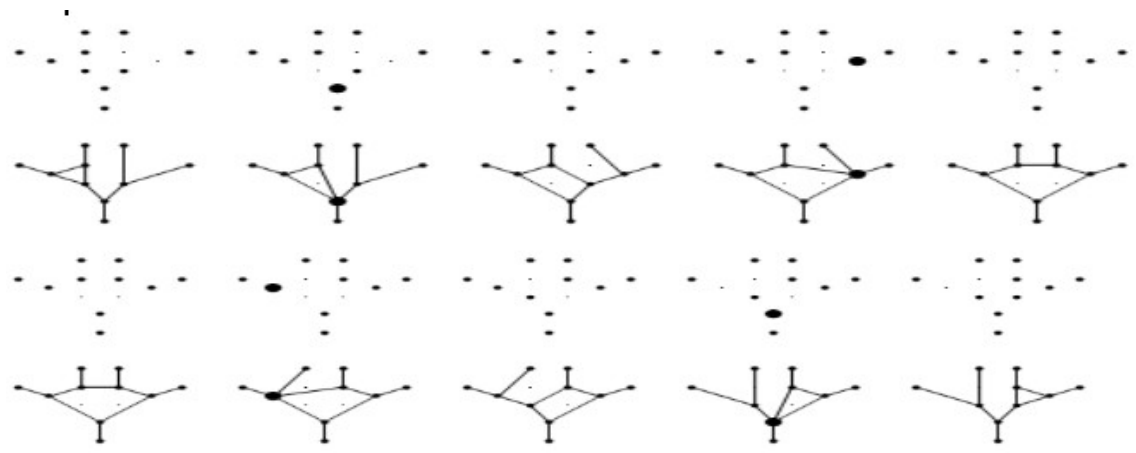

**Figure 10.** Bit swap under Pachner moves

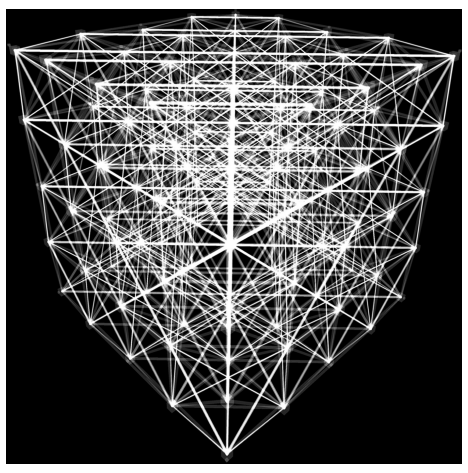

**Figure 11.** 3D cut of a F4 lattice

Under bit swaps, operated by 2-2 Pachner moves (figure 10), spin networks informational content evolves along spin foam history, while geometry (showed in figure 11) remains.

## 4. Topologically encoded tetrad

From trivalent graph to spin network, we must define a tetrad field in su(2) or so(4).

Our graph implicitly encodes quaternions. Hurwitz quaternions integers are the measure in 4D with coordinates on {1,i,j,k} of vectors between one supernode (of figure 11) and any of its 24 neighbors (forming a 24-cell) in D4 sublattice of F4; and with a $\sqrt{2}$ factor and a double $\pi/4$ rotation (in two perpendicular planes, like {1,i} and {j,k}) to the 24 second-neighbors who are the vertices of the 24-cell dual of the previous. The 48-valent supernode is internally designed as a trivalent graph represented on figure 12, a triple binary tree around a central triangle, having 3*8 external leaves, each to be connected to two other leaves of neighbor supernodes. This supernode is also populated with 0 and 1, where 1 is represented by a big dot in figure 12 holding a su(2) value u/3, u/6, i/4, j/4 or i/8 where u is defined as (i+j+k)/ √3.

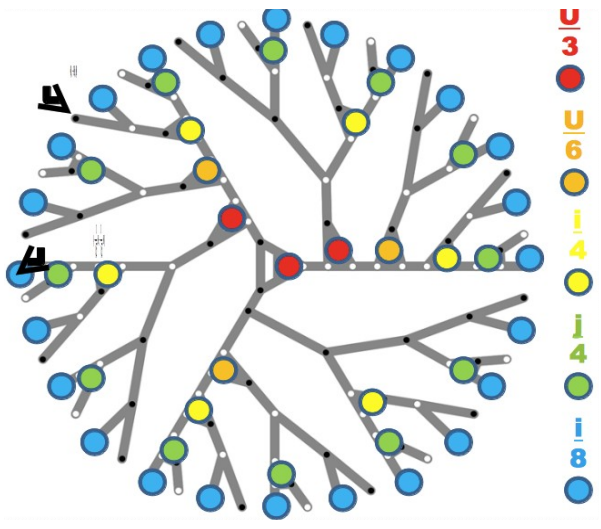

TABLE I
VALUES OF E($n_k$) FIELD

| $K$ | $\{\zeta(K,\Omega_6^{\dagger})\}$ | $\omega_k$ | $\{u_k\}$ | $K$ | $\{\zeta(K,\Omega_6^{\dagger})\}$ | $\omega_k$ | $\{u_k\}$ |
|---|---|---|---|---|---|---|---|
| 1 | $\{1,0,0,0\}$ | $0$ | $\{0,0,0,0\}$ | 25 | $\{-1,0,0,0\}$ | $\frac{1}{2}$ | $\{0,\frac{1}{2\sqrt3},\frac{1}{2\sqrt3},\frac{1}{2\sqrt3}\}$ |
| 2 | $\{\frac{1}{\sqrt2},\frac{1}{\sqrt2},0,0\}$ | $\frac{1}{8}$ | $\{0,\frac{1}{8},0,0\}$ | 26 | $\{-\frac{1}{\sqrt2},-\frac{1}{\sqrt2},0,0\}$ | $\frac{3}{8}$ | $\{0,-\frac{3}{8},0,0\}$ |
| 3 | $\{0,0,1,0\}$ | $\frac{1}{4}$ | $\{0,0,\frac{1}{4},0\}$ | 27 | $\{0,0,-1,0\}$ | $\frac{1}{4}$ | $\{0,0,-\frac{1}{4},0\}$ |
| 4 | $\{0,0,\frac{1}{\sqrt2},-\frac{1}{\sqrt2}\}$ | $\frac{1}{4}$ | $\{0,0,\frac{1}{4\sqrt2},-\frac{1}{4\sqrt2}\}$ | 28 | $\{0,0,-\frac{1}{\sqrt2},\frac{1}{\sqrt2}\}$ | $\frac{1}{4}$ | $\{0,0,-\frac{1}{4\sqrt2},\frac{1}{4\sqrt2}\}$ |
| 5 | $\{0,1,0,0\}$ | $\frac{1}{4}$ | $\{0,\frac{1}{4},0,0\}$ | 29 | $\{0,-1,0,0\}$ | $\frac{1}{4}$ | $\{0,-\frac{1}{4},0,0\}$ |
| 6 | $\{-\frac{1}{\sqrt2},\frac{1}{\sqrt2},0,0\}$ | $\frac{3}{8}$ | $\{0,\frac{3}{8},0,0\}$ | 30 | $\{\frac{1}{\sqrt2},-\frac{1}{\sqrt2},0,0\}$ | $\frac{1}{8}$ | $\{0,-\frac{1}{8},0,0\}$ |
| 7 | $\{0,0,0,1\}$ | $\frac{1}{4}$ | $\{0,0,0,\frac{1}{4}\}$ | 31 | $\{0,0,0,-1\}$ | $\frac{1}{4}$ | $\{0,0,0,-\frac{1}{4}\}$ |
| 8 | $\{0,0,\frac{1}{\sqrt2},\frac{1}{\sqrt2}\}$ | $\frac{1}{4}$ | $\{0,0,\frac{1}{4\sqrt2},\frac{1}{4\sqrt2}\}$ | 32 | $\{0,0,-\frac{1}{\sqrt2},-\frac{1}{\sqrt2}\}$ | $\frac{1}{4}$ | $\{0,0,-\frac{1}{4\sqrt2},-\frac{1}{4\sqrt2}\}$ |
| 9 | $\{\frac{1}{2},\frac{1}{2},\frac{1}{2},\frac{1}{2}\}$ | $\frac{1}{6}$ | $\{0,\frac{1}{6\sqrt3},\frac{1}{6\sqrt3},\frac{1}{6\sqrt3}\}$ | 49 | $\{-\frac{1}{2},-\frac{1}{2},-\frac{1}{2},-\frac{1}{2}\}$ | $\frac{1}{3}$ | $\{0,-\frac{1}{3\sqrt3},-\frac{1}{3\sqrt3},-\frac{1}{3\sqrt3}\}$ |
| 10 | $\{0,\frac{1}{\sqrt2},\frac{1}{\sqrt2},0\}$ | $\frac{1}{4}$ | $\{0,\frac{1}{4\sqrt2},\frac{1}{4\sqrt2},0\}$ | 50 | $\{0,-\frac{1}{\sqrt2},-\frac{1}{\sqrt2},0\}$ | $\frac{1}{4}$ | $\{0,-\frac{1}{4\sqrt2},-\frac{1}{4\sqrt2},0\}$ |
| 11 | $\{-\frac{1}{2},-\frac{1}{2},\frac{1}{2},\frac{1}{2}\}$ | $\frac{1}{3}$ | $\{0,-\frac{1}{3\sqrt3},\frac{1}{3\sqrt3},\frac{1}{3\sqrt3}\}$ | 51 | $\{\frac{1}{2},\frac{1}{2},-\frac{1}{2},-\frac{1}{2}\}$ | $\frac{1}{6}$ | $\{0,\frac{1}{6\sqrt3},-\frac{1}{6\sqrt3},-\frac{1}{6\sqrt3}\}$ |
| 12 | $\{0,-\frac{1}{\sqrt2},\frac{1}{\sqrt2},0\}$ | $\frac{1}{4}$ | $\{0,-\frac{1}{4\sqrt2},\frac{1}{4\sqrt2},0\}$ | 52 | $\{0,\frac{1}{\sqrt2},-\frac{1}{\sqrt2},0\}$ | $\frac{1}{4}$ | $\{0,\frac{1}{4\sqrt2},-\frac{1}{4\sqrt2},0\}$ |
| 13 | $\{-\frac{1}{2},\frac{1}{2},\frac{1}{2},-\frac{1}{2}\}$ | $\frac{1}{3}$ | $\{0,\frac{1}{3\sqrt3},\frac{1}{3\sqrt3},-\frac{1}{3\sqrt3}\}$ | 53 | $\{\frac{1}{2},-\frac{1}{2},-\frac{1}{2},\frac{1}{2}\}$ | $\frac{1}{6}$ | $\{0,-\frac{1}{6\sqrt3},-\frac{1}{6\sqrt3},\frac{1}{6\sqrt3}\}$ |
| 14 | $\{-\frac{1}{\sqrt2},0,0,-\frac{1}{\sqrt2}\}$ | $\frac{3}{8}$ | $\{0,0,0,-\frac{3}{8}\}$ | 54 | $\{\frac{1}{\sqrt2},0,0,\frac{1}{\sqrt2}\}$ | $\frac{1}{8}$ | $\{0,0,0,\frac{1}{8}\}$ |
| 15 | $\{-\frac{1}{2},\frac{1}{2},-\frac{1}{2},\frac{1}{2}\}$ | $\frac{1}{3}$ | $\{0,\frac{1}{3\sqrt3},-\frac{1}{3\sqrt3},\frac{1}{3\sqrt3}\}$ | 55 | $\{\frac{1}{2},-\frac{1}{2},\frac{1}{2},-\frac{1}{2}\}$ | $\frac{1}{6}$ | $\{0,-\frac{1}{6\sqrt3},\frac{1}{6\sqrt3},-\frac{1}{6\sqrt3}\}$ |
| 16 | $\{-\frac{1}{\sqrt2},0,0,\frac{1}{\sqrt2}\}$ | $\frac{3}{8}$ | $\{0,0,0,\frac{3}{8}\}$ | 56 | $\{\frac{1}{\sqrt2},0,0,-\frac{1}{\sqrt2}\}$ | $\frac{1}{8}$ | $\{0,0,0,-\frac{1}{8}\}$ |
| 17 | $\{-\frac{1}{2},\frac{1}{2},\frac{1}{2},\frac{1}{2}\}$ | $\frac{1}{3}$ | $\{0,\frac{1}{3\sqrt3},\frac{1}{3\sqrt3},\frac{1}{3\sqrt3}\}$ | 57 | $\{\frac{1}{2},-\frac{1}{2},-\frac{1}{2},-\frac{1}{2}\}$ | $\frac{1}{6}$ | $\{0,-\frac{1}{6\sqrt3},-\frac{1}{6\sqrt3},-\frac{1}{6\sqrt3}\}$ |
| 18 | $\{-\frac{1}{\sqrt2},0,\frac{1}{\sqrt2},0\}$ | $\frac{3}{8}$ | $\{0,0,\frac{3}{8},0\}$ | 58 | $\{\frac{1}{\sqrt2},0,-\frac{1}{\sqrt2},0\}$ | $\frac{1}{8}$ | $\{0,0,-\frac{1}{8},0\}$ |
| 19 | $\{-\frac{1}{2},-\frac{1}{2},-\frac{1}{2},\frac{1}{2}\}$ | $\frac{1}{3}$ | $\{0,-\frac{1}{3\sqrt3},-\frac{1}{3\sqrt3},\frac{1}{3\sqrt3}\}$ | 59 | $\{\frac{1}{2},\frac{1}{2},\frac{1}{2},-\frac{1}{2}\}$ | $\frac{1}{6}$ | $\{0,\frac{1}{6\sqrt3},\frac{1}{6\sqrt3},-\frac{1}{6\sqrt3}\}$ |
| 20 | $\{0,-\frac{1}{\sqrt2},0,\frac{1}{\sqrt2}\}$ | $\frac{1}{4}$ | $\{0,-\frac{1}{4\sqrt2},0,\frac{1}{4\sqrt2}\}$ | 60 | $\{0,\frac{1}{\sqrt2},0,-\frac{1}{\sqrt2}\}$ | $\frac{1}{4}$ | $\{0,\frac{1}{4\sqrt2},0,-\frac{1}{4\sqrt2}\}$ |
| 21 | $\{-\frac{1}{2},-\frac{1}{2},\frac{1}{2},-\frac{1}{2}\}$ | $\frac{1}{3}$ | $\{0,-\frac{1}{3\sqrt3},\frac{1}{3\sqrt3},-\frac{1}{3\sqrt3}\}$ | 61 | $\{\frac{1}{2},\frac{1}{2},-\frac{1}{2},\frac{1}{2}\}$ | $\frac{1}{6}$ | $\{0,\frac{1}{6\sqrt3},-\frac{1}{6\sqrt3},\frac{1}{6\sqrt3}\}$ |
| 22 | $\{0,-\frac{1}{\sqrt2},0,-\frac{1}{\sqrt2}\}$ | $\frac{1}{4}$ | $\{0,-\frac{1}{4\sqrt2},0,-\frac{1}{4\sqrt2}\}$ | 62 | $\{0,\frac{1}{\sqrt2},0,\frac{1}{\sqrt2}\}$ | $\frac{1}{4}$ | $\{0,\frac{1}{4\sqrt2},0,\frac{1}{4\sqrt2}\}$ |
| 23 | $\{-\frac{1}{2},\frac{1}{2},-\frac{1}{2},-\frac{1}{2}\}$ | $\frac{1}{3}$ | $\{0,\frac{1}{3\sqrt3},-\frac{1}{3\sqrt3},-\frac{1}{3\sqrt3}\}$ | 63 | $\{\frac{1}{2},-\frac{1}{2},\frac{1}{2},\frac{1}{2}\}$ | $\frac{1}{6}$ | $\{0,-\frac{1}{6\sqrt3},\frac{1}{6\sqrt3},\frac{1}{6\sqrt3}\}$ |
| 24 | $\{-\frac{1}{\sqrt2},0,-\frac{1}{\sqrt2},0\}$ | $\frac{3}{8}$ | $\{0,0,-\frac{3}{8},0\}$ | 64 | $\{\frac{1}{\sqrt2},0,\frac{1}{\sqrt2},0\}$ | $\frac{1}{8}$ | $\{0,0,\frac{1}{8},0\}$ |

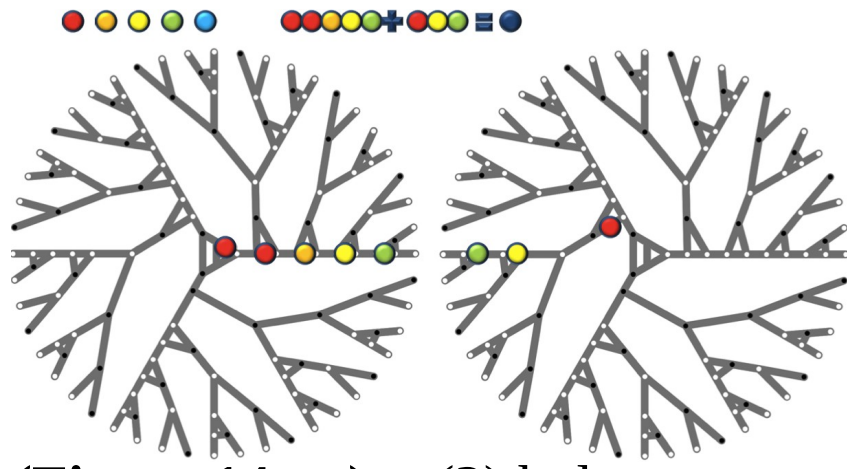

**(Figure 14 - a)** su(2) holonomy

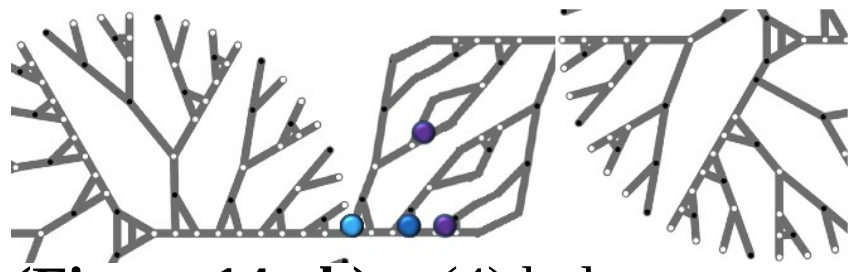

**(Figure 14 - b)** so(4) holonomy

**Figure 12.** Internal supernode design, and its su(2) implicit field

**Figure 13.** su(2) holonomies at 48 leaves

**Figure 14.** opposite (a) su(2) or (b) so(4) (through super-links) holonomies at joining leaves

4.1. su(2) emerging from the only self-dual exceptional polytope, the icositetrachoron, or 24-cell

We take imaginary quaternions as su(2) Lie algebra, unitary quaternions as SU(2) Lie group.

24-cell, radius 2, vertices are integer, twice-unit quaternions: {Permutations(±2,0,0,0), Permutations(±1, ±1, ±1, ±1)}.

24-cell dual, radius √8, vertices are integer, twice-unit quaternions multiplied by (1+i): { Permutations(±2, ±2, 0, 0)}.

D4 is the sublattice of integer lattice having all coordinates of same parity, linked to 24 neighbors at distance 2, and F4 is D4 with additional links to 24 neighbors at distance √8.

4.2. 4D regular space-filling, sphere packing

24-cell, radius 1, centered at two linked nodes of D4 intersect in an octahedron.

24-cell, radius 1, centered on D4 nodes is a 4D space filling.

Radius 1 hyperspheres $S^3$ centered on D4 nodes is an optimal hypersphere packing.

4.3. Supernode as dual of simplicial decomposition

24-cell = 24 octahedrons = 24*(4 or 8)= 96 tetrahedrons to 192 tetrahedrons, because an octahedron can either be split into 8 tetrahedrons, or along 3 different axis into 4 tetrahedrons.

Dual 2-complex of 4D Space filling by 24-cells centered on D4 lattice nodes can holds 2 bits by octahedron, defining the chosen decomposition, and will be studied in a further work.

4.4. Supernode as triple-tree

In D4 each supernode is 24-valent, and can be extruded to a triple tree of 24 trivalent nodes; in F4 24 has to be replaced by 48.

A natural replacement of half of the nodes by a triangle encode in harmony 0 and 1 in the tree, so that the path from the center to each of 48 leaves is a binary code. We finally get 144 trivalent nodes.

4.5. Supernode interconnection

Topological encoding labels each of 48 leaves.

Transitive valuation of each tree-level from the root to any leaf defines a quaternion field over the nodes and edges of the supernode, as SU(2) holonomy along the spin network inside the supernode.

For the $K^{st}$ leaf, where $K$ has binary decomposition $K=\Sigma b_{n,K}\, 2^n$, the SU(2) holonomy is given [2] by:

$$\zeta(K) \coloneqq ((((\exp(b_{5,K}\frac{2\pi}{3}\mathbb{u})\exp(b_{4,K}\frac{2\pi}{3}\mathbb{u}))\exp(b_{3,K}\frac{2\pi}{6}\mathbb{u}))\exp(b_{2,K}\frac{2\pi}{4}\mathbb{i}))\exp(b_{1,K}\frac{2\pi}{4}\mathbb{j}))\exp(b_{0,K}\frac{2\pi}{8}\mathbb{i}) \quad (1)$$

The su(2) imaginary quaternion is its Log, product of an angle $\omega_K$ by an unitary imaginary direction $u_K$, such that $\zeta(K)=\exp(2\pi\ \omega_K\ u_K)$, as given in table from figure 13.

A module of 2 or 8 depending on $b_0$, and this quaternionic argument, define the translation between connected supernodes, encoding spacetime (figure 14).

The tetrad $e$ should be deduced from the connection $\omega$ by reverting the well-known [3] equation:

$$\omega[e]_{\mu}^{IJ}=2\ e^{\nu[I}e_{[\nu;\mu]}^{\ \ J]}+e_{\mu K}e^{\nu I}e^{\sigma J}e_{[\nu;\sigma]}^{\ \ K} \qquad (2)$$

The mean tetrad $e$ for the supernode in this pure crystal configuration is by symmetry $e$={1,i,j,k}, giving a metric tensor of Minkowski flat space-time, while some bit-swapping in the supernode, encoding particles as in the next paragraph, will break the symmetry and induce a curvature.

Therefore, E.F.E. without matter holds, and E.F.E. with matter can be computed.

## 5. Bosons and fermions

The swapping of two bits (a one and a zero) from the default configuration encodes a bitswap.

Twisted e8 basis: {$\omega_L$/4, W/4, B/4, $\omega_R$/4,k,y,m,c}={higgsonic part, gluonic part}

### 5.1. Electric charge

If<o.( W/4+ B/4)>/2 is even: <o. (W/4+ B/4 +k)>/4 - <o.(y+m+c)>/12

If<o.( W/4+ B/4)>/2 is odd: - <o. ( B/4+ $\omega_R$/4 +k)>/4 - <o.(y+m+c)>/12

### 5.2. Color charge

Trivial from <o.y>,<o.m>,<o.c> with –y=b, -m=g, -c=r

### 5.3. Example: the blue up quark

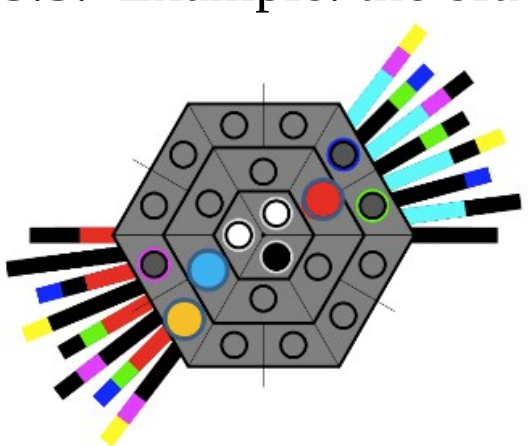

**Figure 15.** Encoding blue up quark on 21 bitswaps

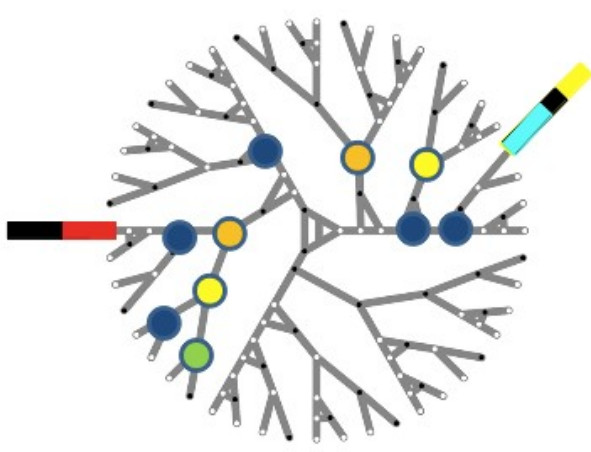

**Figure 16.** Bitswaps at dots, higgsonic & gluonic leaves

| | | | |
|---|---|---|---|
| {-2, 0, -2, 0, -2, 0, 0, 0} | 1 | | Positron |
| {-2, 0, -2, 0, 0, -2, 0, 0} | 2/3 | | Up |
| {-2, 0, -2, 0, 0, 0, -2, 0} | 2/3 | | Up |
| {-2, 0, -2, 0, 0, 0, 0, -2} | 2/3 | | Up |
| {-2, 0, -2, 0, 0, 0, 0, 2} | 1/3 | | Anti Down |
| {-2, 0, -2, 0, 0, 0, 2, 0} | 1/3 | | Anti Down |
| {-2, 0, -2, 0, 0, 2, 0, 0} | 1/3 | | Anti Down |

**Figure 17.** Coordinates as e8 root of one blue up quark

Figure 15 shows how 3 central bit swaps encode fermion/boson state and family, then 3 other define higgsonic part, and the 3 opposite define gluonic part.

In a more natural encryption scheme, but less economical, inter-supernodes links are replaced by superlinks, with superlinks loop edges as imaginary octonions and SU(2) replaced by E8 holonomies, will define 192 fermions in superlinks and 48 bosons at supernodes leaves.

## 6. Conclusion

We described a gravitational tetrad implicit from topological information, with:

- su(2) implicit algebra whose holonomy gives a translation operator
- e8 implicit algebra whose holonomy defines E8 roots, as quantum numbers of 48 bosons and 192 fermions in {$\omega_L$/4, W/4, B/4, $\omega_R$/4,k,y,m,c}
- Implicit trace of $\omega$ connection. Tetrad $e$ defined from tr($\omega$)
- 4D space triangulated along 4D space filling by 24-cell, or trivalent informed supernodes
- pentavalent spin foam of history of Pachner 2-2 moves on trivalent spin network.

Thanks to encouraging share of thoughts with the quantum gravity community, this work is refining.